# Parameter dependence of resonant spin torque magnetization reversal

*L. Fricke, S. Serrano-Guisan and H. W. Schumacher*

Physikalisch-Technische Bundesanstalt, Bundesallee 100, D-38116 Braunschweig, Germany.

Corresponding author:
Hans Werner Schumacher

E-mail: hans.w.schumacher@ptb.de,
phone: +49 (0)531 592 2500
fax: +49 (0)531 592 69 2500

## Abstract

We numerically study ultra fast resonant spin torque (ST) magnetization reversal in magnetic tunnelling junctions (MTJ) driven by current pulses having a direct current (DC) and a resonant alternating current (AC) component. The precessional ST dynamics of the single domain MTJ free layer cell are modelled in the macro spin approximation. The energy efficiency, reversal time, and reversal reliability are investigated under variation of pulse parameters like direct and AC current amplitude, AC frequency and AC phase. We find a range of AC and direct current amplitudes where robust resonant ST reversal is obtained with faster switching time and reduced energy consumption per pulse compared to purely direct current ST reversal. However for a certain range of AC and direct current amplitudes a strong dependence of the reversal properties on AC frequency and phase is found. Such regions of unreliable reversal must be avoided for ST memory applications.

Article

## I. INTRODUCTION

Spin transfer torque [1,2] magnetization reversal of the free layer of a magnetic tunnel junction is a promising method for writing information in non-volatile scalable magnetic random access memories (MRAM) [3]. However for future high density ST-MRAM, a reduction of switching power and critical current density $j_C$ for magnetization reversal is required. Here, resonant spin torque magnetization reversal driven both by direct currents (DC) and superimposed resonant alternating currents (AC) [4-7] or direct currents combined with alternating fields [8] might allow a more efficient spin torque magnetization reversal compared to ST reversal purely by DC currents. For applications of such resonant ST reversal schemes in future magnetic random access memories (MRAM) the reversal reliability under parameter stray is of crucial importance. So far different approaches towards resonant ST reversal using combined DC and AC currents have been proposed [4-7]. Cui et al. have studied reversal by an initial AC pump excitation followed by a DC pulse [6] and observed a strong influence of the phase of the resonant AC excitation at the onset of the DC pulse on the switching probability. Florez et al. studied ST switching by a current pulse assisted by a continuously applied resonant AC excitation [5,7]. Here the phase relation was not systematically varied and the observed frequency dependent reversal was assigned to frequency locking of preswitching precessional modes to the AC excitation. Rivkin and Ketterson [4] showed that ST reversal can be optimized when using a series of *chirped* current pulses instead of a single direct current pulse. Here the series of ultra short rectangular current pulses is tailored such that it tracks the time dependency of the resonance frequency during ST reversal leading to fast and highly efficient reversal. However the realization of an optimum chirp for *each* memory cell of a ST-MRAM is demanding especially when considering a certain parameter stray over the cell array.

Here, we consider a similar but simpler to implement approach to resonant ST reversal: application of a direct current pulse and a superimposed pulse of a resonant alternating current of constant frequency. ST magnetization reversal as a function of AC and DC current amplitudes is simulated in a macrospin model. To investigate the reversal reliability under parameter stray the AC frequency and phase is varied. We find an optimum range of AC and direct current amplitudes where robust resonant ST reversal is obtained with faster switching time and reduced energy consumption per pulse compared to purely direct current ST reversal. However for a certain range of AC and direct current amplitudes a strong dependence of the reversal properties on AC frequency and phase is found. These regions of unreliable reversal are not suitable for ST-MRAM applications.

## II. MODEL

To model the dynamics of resonant ST magnetization reversal we consider an elliptic magnetic tunnelling junction (MTJ) stack having 285 x 140 $nm^2$ lateral dimensions as sketched in Fig. 1(b). The simplified MTJ stack consists of a CoFeB (2 nm) free layer separated from the fixed CoFeB (2 nm) reference layer by a 2 nm insulating layer. For the free layer no other anisotropy than shape anisotropy is taken into account. To minimize stray field effects from the reference layer on the free layer the reference layer is antiferromagnetically coupled to a second pinned CoFeB layer of 2 nm thickness. Together the reference layer and the pinned layer mimic the usual synthetic antiferromag-netic pinned layer structures used in ST-MRAM MTJ stacks, today. The exchange field of the antiferromagnetic coupling is set to - 200 mT. Additionally, the orientation of the magnetization of the pinned layer is fixed by a bias field of 300 mT. These values of exchange field and bias field assure a negligible magnetization motion of pinned and reference layer during the simulations. For all CoFeB layers we assume a saturation magnetization of $M_s$ = 1.8 T, a Gilbert damping parameter of 0.01, and a spin polarization of 0.6. These parameters are comparable to values measured in real CoFeB/MgO based magnetic tunnelling junctions [9,10]. As sketched in Fig. 1 (b) the orientation of the pinned layer bias field and hence the orientation of the magnetization of the antiferromagnetically coupled reference layer are tilted by θ = 6° in plane with respect to the free layer easy axis. Such small tilts of θ = 2…10° between free and reference layer allows for optimized ST magnetization reversal with minimum critical current densities $j_C$ when simulating direct current ST reversal.

Simulations of the devices were carried out using a commercial magnetization dynamics simulation software package [11]. In the simulations the Landau-Lifshitz-Gilbert equation [12] including the Slonczewski ST term [1] was solved including the abovementioned anisotropies, inter layer couplings and bias fields. Simulations were done in the macrospin approximation i.e. by considering a homogeneously oriented magnetization in each layer. No thermal excitations were taken into account.

Simulations of DC ST reversal delivered a critical DC current density of $j_C = 3.73 \cdot 10^6$ A/cm$^2$ comparable to critical current densities found experimentally [5-7,9,10]. The magnetization switching time $t_{sw}$ at the given $j_C$ is measured between the inset of the DC cur-rent and the moment the easy axis component $m_X$ of the free layer magnetization M overcomes the hard axis and is found to be $t_{sw} \sim 4.5$ ns.

The pulse shape used in the resonant ST reversal simulations is sketched in Fig. 1(a). 5 ns current density pulses are applied consisting of a square wave component $j_{DC}$ and a superimposed AC component $j_{AC}$. Vanishing rise and fall times of the current pulses are assumed. In the simulations $j_{DC}$ is varied in the range of 0.7 … 1.5 $j_C$ and $j_{AC}$ in the range of 0 … 2.75 $j_C$. To test the reversal reliability under parameter stray both the AC frequency $f$ and the phase $\phi$ of the AC oscillations at the inset of the pulse are varied for each pair of current parameters $j_{AC}$, $j_{DC}$. $f$ is varied around the ferromagnetic resonance frequency of the free layer magnetization $f_{ST}$ = 5.05 GHz in the range of 4.8 to 5.2 GHz in steps of 0.1 GHz. $\varphi$ is varied from 0 to $\pi/2$ in steps of $\pi/12$.

### III. RESULTS AND DISCUSSION

Fig. 2 shows the pulse shape and the time evolution of the easy and hard axis component of the free layer magnetization ($m_X$, $m_Y$, respectively) for two different resonant spin torque pulses. The left hand side of the figure (a)-(c) shows magnetization reversal by a typical resonant ST pulse. (a) shows the time dependent pulse shape. Here, $j_{DC}$ = 1.35 $j_C$, $j_{AC}$ = 2.43 $j_C$, $f$ = 4.9 GHz, and $\varphi$ = 0. (b) and (c) show the time dependent easy axis and hard axis components of M $m_X$ (b) and $m_Y$ (c), respectively. For the given current polarity the applied ST pulse leads to a negative effective damping of the magnetization of the free layer. Hence the pulse excites a precessional motion of M with steadily increasing in plane angular excursion. At $t_{SW}$ = 1.1 ns (dashed vertical line) M overcomes the hard axis and magnetization reversal takes place. After $m_X$ has changed sign ST enhances the in-trinsic Gilbert damping and M relaxes to the reversed easy axis orientation by damped precession.

Fig. 3 shows a grey scale map of the reversal probability $P_{REV}$ under parameter stray as function of $j_{AC}, j_{DC}$. For every current parameter pair the reversal probability under parameter stray is computed in the following way. First the number of parameter sets $(f,\phi)$ for the given pair of current values $(j_{DC}, j_{AC})$ which lead to magnetization reversals is counted. Then the number is normalized by the total number of parameter sets $(f,\phi)$ for the given current value. Fully reliable ST reversal corresponds to $P_{REV} = 1$. Here, reversal is induced for every set of $(f,\phi)$ during application of a 5 ns pulse of the given $(j_{DC}, j_{AC})$. In Fig. 3 fully reliable reversal ($P_{REV}=1$) is plotted white whereas no reversal ($P_{REV}=0$) is plotted black. Intermediate values corresponding to unreliable reversal are plotted in grey shades.

In the map four regions having characteristic reversal properties (1)-(4) are found. For small amplitudes of $j_{DC}, j_{AC}$ ((1), black) the applied pulses are too weak to induce magnetization reversal and no reversal is found. In contrast, for large values of $j_{DC}, j_{AC}$ ((2), white region) fully reliable reversal is found regardless of $f$ and $\varphi$. Here, a simulated parameter stray of the AC current does not affect the reliability of the ST reversal. Near the boundary between reversal and no reversal ((3), grey shades) the reversal is not fully reliable. Here the reversal properties depend on $f$ and $\varphi$, so that for some specific parameter settings the magnetization reverses whereas for others not. Such unreliable reversal near the reversal boundary can be expected. During pulse application precessional oscillations with increasing amplitude are excited. Near the reversal boundary M overcomes the hard axis approximately at the moment of pulse decay i.e. after about 5 ns. Slight parameters changes may now slightly delay the reversal and the pulse decays before magnetization reversal has taken place thus resulting in unreliable reversal.

More surprisingly an additional region of reduced reversal reliability (4) is found in the upper left of the plot. Note that in this region unreliable reversal occurs for direct current components *above* the critical switching threshold $j_{DC} > j_C$. The existence of this region can not be explained in the same picture as region (3). Here, unreliable reversal is due to the development of steady state precessional oscillations which are phase locked to the AC current excitation. This region has no well defined boundary but rather appears as a set of unconnected spots. This becomes more clear from the inset to Fig. 3. In the inset the region marked by the black rectangle in the main figure is plotted. The data in the inset is based on a simulation with higher resolution of the $j_{DC}, j_{AC}$ mesh. Also in the higher resolution simulation unconnected areas of reduced reversal probability appear. This reflects the strong nonlinearity of the ST magnetization dynamics. Only for very specific initial conditions of phase and frequency phase locked steady state oscillations are stabilized whereas for slightly different initial conditions fast magnetization reversal may occur. Furthermore, steady state precessional oscillation are found for different values of AC phase and frequency $(f,\phi)$ for the different current parameter pairs $(j_{DC}, j_{AC})$. This strong dependence of the reversal properties on the exact pulse parameters makes this region unsuitable for applications in ST-MRAM.

The occurrence of phase locked steady state oscillations can be well observed in Fig. 2. The right hand side (d)-(f) shows an example of the applied pulse and the time evolution of $m_X$, $m_Y$ from region (4) of Fig. 3. The parameters of the pulse in (d) are $j_{DC} = 1.05\ j_C, j_{AC} = 0.14\ j_C, f = 5.1$ GHz, $\varphi = 0$. As seen both in $m_X$ (e) and $m_Y$ (f) M is first excited to precession with steadily increasing amplitude. However at t = 3.2 ns (vertical dashed line) a steady oscillation develops. Here the precessional frequency is locked to the excitation frequency $f$ and M oscillates with constant amplitude until the decay of the current pulse. Then M relaxes back to the initial easy axis orientation by damped precession. Note that this phase locked resonant oscillation inhibiting reliable ST reversal can only develop under the influence of the applied AC current component. Fig. 2(e) also shows the reversal trajectory of $m_X$ for a DC current pulse of same amplitude $j_{DC}$ = 1.05 (grey curve) but without the AC component (i.e. $j_{AC}$ = 0). Here, in contrast, $j_{DC}$ induces ST reversal within about 4 ns.

Note that already in the rather simple macro spin model the strongly nonlinear ST dynamics can induce phase locked steady state oscillations and hence unreliable reversal. Although the exact parameter dependence will certainly differ similar effects could be expected in more realistic micromagnetic simulations. These are however beyond the scope of the present work.

In Fig. 3 also lines of constant electrical power $P_{el}$ of the applied pulses are shown (grey lines). The applied power is normalized to the power $P_0$ of a 5 ns direct current pulse of $j_C$. The operation range of a realistic ST-MRAM including a reliability margin might be chosen in the range of about 1.2 … 2 $P_0$. Here, outside the region (4) of unreliable reversal, a wide range of $j_{DC}, j_{AC}$ is found where reliable ST reversal takes place.

As seen from Fig. 3 the application of a resonant AC current component during ST reversal as studied here does not allow a *reduction* of peak current density or pulse power compared to direct current ST reversal. However it might allow faster MRAM operation and reversal by pulses with lower total energy $E_{el}$ as shown in Fig. 4.

Fig. 4 shows a grey scale map of the maximum of the magnetization reversal time $t_{sw,max}$ found for every pair of current parameters $j_{DC}, j_{AC}$ under variation of $f$ and $\phi$. Here, the maximum of $t_{sw}$ is considered to model the worst case which can occur in a ST-MRAM array under parameter stray. Also in this plot the same four regions (marked by numbers (1)-(4) as discussed before) can be identified. Again, in region ((1), white) no reversal takes place. In the unreliable regions (3), (4) the maximum reversal times are close to the maximum value of 5 ns given by the pulse duration. In region (2) $t_{sw,max}$ gradually decreases with increasing $j_{DC}, j_{AC}$ and fast ST reversal takes place. Here, $t_{sw,max}$ also determines the *minimum* pulse duration $T_{P,min}$ for the given current parameters $j_{DC}$, $j_{AC}$ that still induces *reliable* ST magnetization reversal. Thus in the regions of reliable reversal the duration of the applied pulse could be reduced down to $T_{P,min}$ (plus an appropriate operation margin) thereby reducing the total energy of the applied pulse.

The minimum energy per pulse $E_{el,min}$ can be calculated by assuming a pulse duration of $T_{P,min} = t_{sw,max}$. Again, the minimum energy is normalized to the minimum reversal energy $E_0$ for a DC pulse of $j_C$ and duration $t_{sw}$ that induces magnetization reversal. A combination of reliable reversal as discussed above and low energy per reversal pulse allows most efficient ST-MRAM operation. This optimum operation region (5) is situated left above the region of unreliable switching (4) i.e. above the white tilted line in Fig. 4. In this region the electrical energy per pulse is below 0.8 $E_0$. In the other region of reliable reversal (2) no significant reduction of reversal energy compared to DC pulses is obtained. As a consequence, the application of a rather small AC current component of $j_{AC} < 0.5\ j_{DC}$ on top of the DC current pulse seems to be promising for realization of fast and yet energy efficient future ST-MRAM.

## IV. CONCLUSIONS

We have presented numerical studies of ultra fast resonant spin torque (ST) magnetization reversal in magnetic tunnelling junctions (MTJ) driven by current pulses having a direct and a resonant AC component. Focus was put on the reversal probability under parameter stray of AC frequency and phase. A current parameter range unsuitable for ST-MRAM operation was identified where the reduced reversal probability was linked to the existence of phase locked steady state precession of the magnetization. Furthermore an optimum region of AC and direct current amplitudes was found where robust resonant ST reversal can be obtained with faster switching time and reduced energy consumption per pulse compared to purely direct current ST reversal. Such resonant ST magnetization reversal schemes may improve operation of ultra fast and yet efficient future ST-MRAM devices.

### ACKNOWLEDGEMENT

The research leading to these results has received funding from the European Community's Seventh Framework Programme, ERA-NET Plus, under Grant Agreement No. 217257 within the Joint Research Project Nanospin.

**REFERENCES:**

[1]  J. C. Slonczewski, "Current-driven excitation of magnetic multilayers" J. Magn. Magn. Mater. **159**, L1-L7 (1996).

[2]  L. Berger, "Emission of spin waves by a magnetic multilayer traversed by a current" Phys. Rev. B **54**, 9353-9358 (1996).

[3]  Yiming Huai, Frank Albert, Paul Nguyen, Mahendra Pakala, and Thierry Valet, "Observation of spin-transfer switching in deep submicron-sized and low-resistance magnetic tunnel junctions", Appl. Phys. Lett. **84**, 3118-3120 (2004).

[4]  K. Rivkin and J.B. Ketterson, "Switching spin valves using rf currents" Appl. Phys. Lett. **88**, 192515 (2006).

[5]  S. H. Florez J. A. Katine, M. Carey, L. Folks, and B. D. Terris, "Modification of critical spin torque current induced by rf excitation" J. Appl. Phys. **103**, 07A708 (2008).

[6]  Y-T. Cui, J. C. Sankey, C. Wang, K. V. Thadani, Z.-P. Li, R. A. Buhrman, and D. C. Ralph, "Resonant spin-transfer-driven switching of magnetic devices assisted by microwave current pulses", Phys. Rev. B **77**, 214440 (2008).

[7]  S. H. Florez, J. A. Katine, M. Carey, L. Folks, O. Ozatay, and B. D. Terris, "Effects of radio-frequency current on spin-transfer-torque-induced dynamics" Phys. Rev. B **78**, 184403 (2008).

[8]  Paul. P. Horley, Vitor R. Vieira, Peter M. Gorley, Vitalii K. Dugaev, Jamal Berakdar, and Józef Barnas, "Influence of a periodic magnetic field and spin-polarized current on the magnetic dynamics of a mondomain ferromagnet" Phys. Rev. B **78**, 054417 (2008).

[9]   J. C. Sankey, Y-T. Cui, J. Z. Sun, J. C. Slonczewski, R. A. Buhrman and D. C. Ralph, "Measurement of the spin-transfer-torque vector in magnetic tunnel junctions" Nature Physics **4**, 67 (2007).

[10]  S. Serrano-Guisan, K. Rott, G. Reiss, J. Langer, B. Ocker, H. W. Schumacher, "Biased quasi ballistic spin torque magnetization reversal" Phys. Rev. Lett., **101**, 087201 (2008).

[11]  M. R. Scheinfein, LLG Micromagnetic Simulator, http://llgmicro.home.mindspring.com/

[12]  L. Landau, and E. Lifshitz, Phys. Z. Sowjetunion **8**, 153, (1935); T. L. Gilbert, Phys. Rev. **100**, 1243 (1955).

**FIGURE CAPTIONS:**

**FIGURE 1:**

a) Sketch of the resonant current pulse applied to the cell consisting of a DC current density pulse $j_{DC}$ and an AC current density pulse of amplitude $j_{AC}$. Frequency $f$ and phase $\varphi$ are sketched. b) Sketch of the simulated device. Top: lateral dimensions of the cell and relative orientation of free and reference layer magnetization. Bottom: stack structure of the model MTJ.

**FIGURE 2:**

Resonant ST pulse excitations at (a) $j_{DC} = 1.35\,j_c$, $j_{AC} = 2.43\,j_c$, $f = 4.9$ GHz, $\varphi = 0$ and (d) $j_{DC} = 1.05\,j_c$, $j_{AC} = 0.13\,j_c$, $f = 5.1$ GHz, $\varphi = 0$ and their corresponding time evolution of $m_X$ (b, e) and $m_y$ (c, f). Grey curve in e) shows the time evolution of $m_X$ for a purely direct current pulse of $j = j_{DC} = 1.05\,j_C$.

**FIGURE 3:**

Reversal reliability $P_{REV}$ as a function of normalized DC and AC current densities. Black: no reversal, white: reliable reversal. The inset shows a more detailed simulation of $P_{REV}$ of the region (4) marked by the black rectangle. The gray contour lines indicate constant normalized pulse power $P_{el}$ of the applied current pulses.

**FIGURE 4:**

Grey map of the maximum switching time $t_{SW,max}$ as a function of normalized AC and DC current density. Region (5) above the white tilted line is characterized by a minimum pulse energy $E_{el,min}/E_0$ needed to induce magnetization reversal.

**FIGURES:**

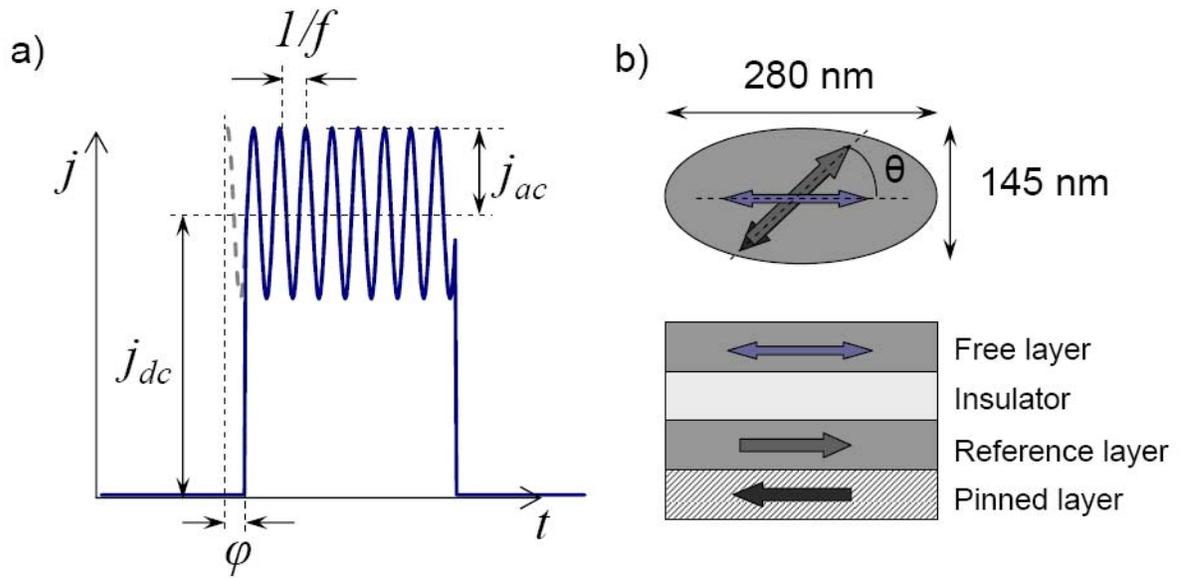

L. Fricke et al,

**Figure 1**

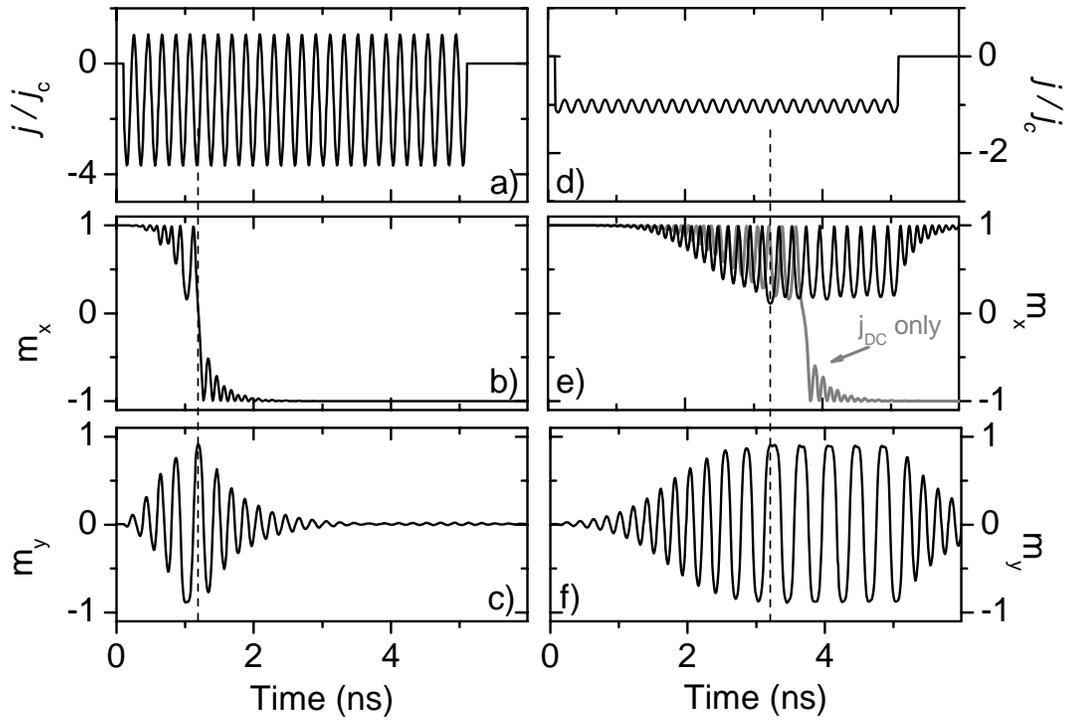

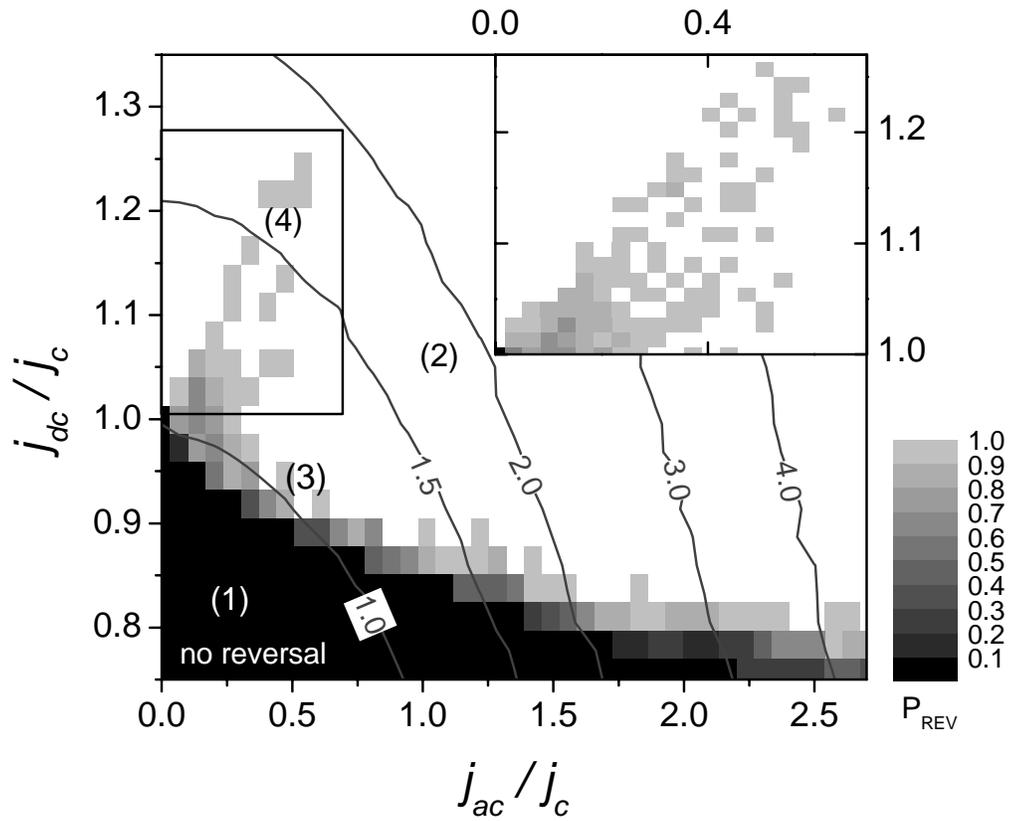

L. Fricke et al,

Figure 3

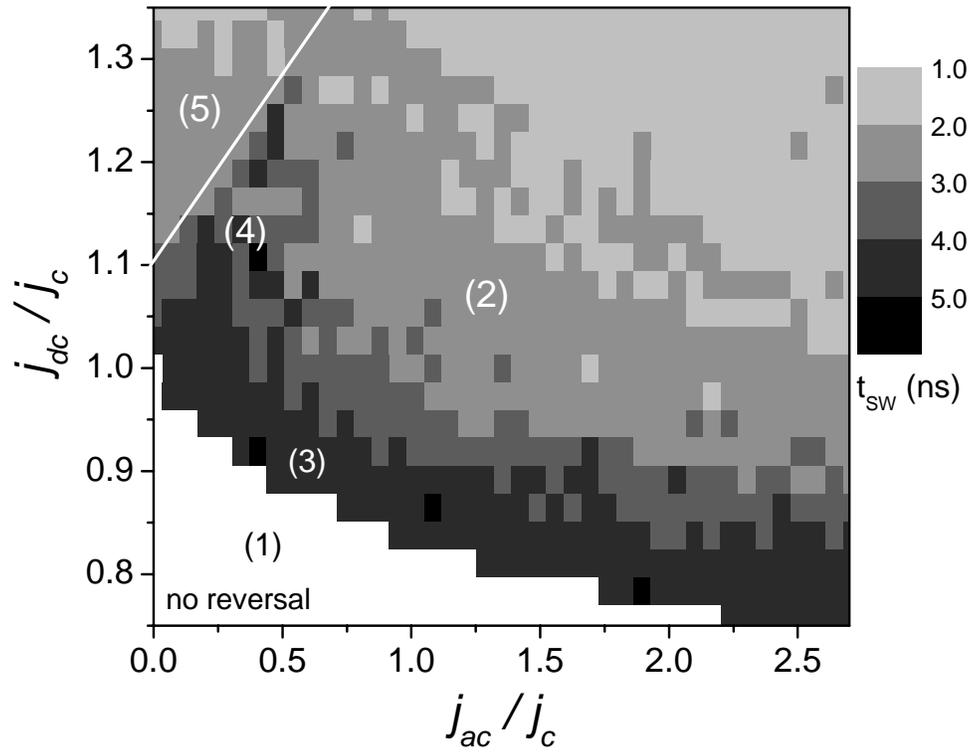

L. Fricke et al,

**Figure 4**